\documentclass[conference]{IEEEtran}

\usepackage{graphicx}
\usepackage[linesnumbered,ruled,vlined]{algorithm2e}
\usepackage{array}
\usepackage{hyperref}
\def\BibTeX{{\rm B\kern-.05em{\sc i\kern-.025em b}\kern-.08em
    T\kern-.1667em\lower.7ex\hbox{E}\kern-.125emX}}
\begin{document}

\title{Implementation of Course Recommender System for Virtual University of Pakistan}

\author{\IEEEauthorblockN{Aleem Akhtar}
\IEEEauthorblockA{\textit{Dept. of Computer Science} \\
\textit{Virtual University of Pakistan}\\
aleem.akhtar@seecs.edu.pk}
}

\maketitle

\begin{abstract}
Universities working in Pakistan are offering a comprehensive set of degree programs for different levels. Virtual University of Pakistan is country's first institution completely based on modern information and communication technologies. It offers education in many different majors and various areas of study are available. Multiple courses are offered in each program that satisfy several general requirements of degree. Selection of courses that align with competency and interest can become an important factor in determining final score (CGPA) of student. For this purpose, a web-based course recommender system specifically designed for courses offered at Virtual University is developed. User-based collaborative filtering and rating-prediction approach is used for calculation of expected marks and grades. System is tested against 470 currently available courses and simulated data of 2600 students. Test results showed that expected marks are somehow dependent on student's average marks in already studied courses and average marks of similar students in target course. Accuracy of implemented system is measured using Mean Absolute Error for 100 observations. MAE value came out to be in acceptable range.  
\end{abstract}

\begin{IEEEkeywords}
Recommender System, Collaborative Filtering, VU
\end{IEEEkeywords}

\section{Introduction}
\label{Introduction}
There are hundreds of both private and public educational institutions working in Pakistan providing education from school level to doctorate level in various fields. Average annual enrollment in Universities of Pakistan is in thousands. Each University offers education in many different majors and various areas of study are available. Multiple courses are offered in each program that satisfy several general requirements of degree in that area. These courses are divided into two categories: \textit{required} courses that are compulsory for student to take and \textit{elective} courses that give student choice of specialization in specific field. Few well-known disciplines offered by Universities in Pakistan are:
\begin{itemize}
	\item Engineering and Technology
	\item Biological and Medical Sciences
	\item Arts \& Humanities
	\item Business and Management Education
	\item Agriculture
	\item Veterinary Sciences
	\item Physical and Social Sciences
\end{itemize}

\subsection{Virtual University of Pakistan}
\label{VUP}
Virtual University (VU) of Pakistan is a public University established in 2002 by Government of Pakistan to promote distance-learning education in modern information and communication system technologies. Virtual University is country's first educational institution completely based on delivering lectures through Internet. The total number of students currently studying is nearly 75,000 with 25,000 new students enrolled in 2019 from Pakistan and other overseas countries. University offers academic degrees in Information Technology, Computer Science, Economics, Business administration, Arts, Education, and Physical sciences with many programs offered in BS, Masters, MS/MPhil, and PhD. Each degree or program is composed of both required and elective courses in every semester as defined by their study schemes \cite{vustudyscheme}.

Degree completion requirement at Virtual University is mainly based on two parameters: minimum number of courses passed (credit hours) and minimum Cumulative Grade Point Average (CGPA). VU follows semester system and students enroll themselves in different courses at the start of each semester through Learning Management System (VU-LMS) \cite{vulms}. Course selection process is quick and easy for those students who had already planned out their courses but for some students it can become difficult to select courses which align with their competency and are relatively easier to earn good grade. Selection of courses which are outside student's interest and are naturally considered difficult to get good grades (as per past records) can harm overall percentage and CGPA of student. Therefore, keeping this issue in mind, a course recommender system for Virtual University is presented in this paper with the aim to help students in course selection process.

Rest of this paper is divided into five sections. The first section outlines related work done in the similar field followed by discussion on recommender systems in the second section. The third and fourth sections provide details of design and implementation of VU-CRS and results discussion respectively. The fifth and the last section concludes the paper.

\section{Related Work}
\label{RelatedWork}
For past many years, recommendation systems in the field of course selection and computer based learning has been topic of interest for many scientific researchers. Techniques and methods for recommendations have been presented and evaluated since the beginning of 1950. In this section, some of the similar work have been discussed.

Rickel \& Johnson, presented selection of course content through adaptive model in 1997 \cite{rickel1997integrating}. In 2012, Lobo and Aher, used combination of Weka and various association rules such as Apriori, Filtered, Tertius, and predictive Apriori to implement a course recommender system for Walchand Institute of Technology, India \cite{aher2012comparative}. Another similar study was carried out by Nan Jiang, who designed and developed a course recommender system for College of Wooster as part of his thesis. Using various collaborative filtering techniques, his system recommends series of courses for subsequent semesters based on courses taken by similar students \cite{jiang2017building}. Chau \textit{et al.} presented a recommender system to assist programming course instructors in preparation of most relevant course content. Structure of course is deduced using programming examples set by course instructor and based on these examples, learning material is recommended \cite{chau2018learning}.

O'Mahony and Barry Smyth published their research on recommender systems for enrollment in online courses. A course recommender system is developed by authors based on online enrollment application for University College of Dublin. In this research, different factors that can impact student course choices are highlighted with solution to many key considerations is also proposed. In this system, core courses are used to recommend elective courses \cite{o2007recommender}. Farzan and Brusilovsky used incentive based technique to increase the interest in submission of course feedback. This technique was adopted through career goal interface in CourseAgent which turned course rating as a part of feedback to track their career goals progress. Interest of students was significantly increased through use of incentive mechanism \cite{farzan2011encouraging}. A similar work based on assessment of graduate attributes was presented by Bakhshinategh \textit{et al.} in 2017. Once course is finished, student rate the improvement in their graduating attributes and courses are suggested based on collaborative filtering algorithms \cite{bakhshinategh2017course}. In 2018, Bridges \textit{et al.} presented a research-oriented study to propose a course recommender system based on grade and enrollment data. Graphical analysis techniques are used to analyze semester courses sequence to generate a balanced course transition graph between current grades, course popularity and expected improvements \cite{bridges2018course}. A course recommendation system based on course selection of peers is presented by Bercovitz \textit{et al.} ``CourseRank: A Social System for Course Planning'' is well appreciated concept presented in this paper. This system is basically a course evaluation and planning social system therefore course recommendation is not very much flexible.

Some other studies are also available that focus on using recommendation techniques to present learning material in adaptive nature. This include adaptive modeling system based on difficulty of learning units by Pask \cite{pask1964adaptive} and adaptive modeling system for sequencing of learning units by Tennyson and Christensen \cite{tennyson1988mais} to increase the interest of students and better understanding of modules. André \textit{et al.} \cite{andre1997webpersona} and Rickel \textit{et al.} \cite{rickel1997integrating} in their respective publications dug deep into this field and introduced artificial intelligence methods to enhance adaptive methods for computer based learning. Using concepts introduced by these researchers, many application systems are implemented to cover this area. The ELM-ART II, InterBook and AHA! are such systems that use adaptive techniques to create and present adaptive learning materials.

There are many other similar studies published in the field of course recommender systems. Each system uses one or more attributes such as past performance, career goal, graduate attributes, course rank, or student feedback to either recommend a course or course content. It is not possible to review each system and it requires a separate study targeting only literature review of course recommendation systems. However, in this paper, similar techniques as discussed in cited papers is used to design and implement course recommendation system for Virtual University of Pakistan.

\section{Recommender Systems}
\label{RS}
In the book ``\textit{The Long Tail}'', Chris Anderson said: ``\textit{We are leaving the age of information and entering the age of recommendation}'' \cite{anderson2006long}. An overwhelming amount of information is surrounding people which helps them in making better decisions. However, quality of such decisions get reduced when same information is overloaded. Internet particularly depicts this phenomenon. In one minute, more than 400 hours of YouTube videos are uploaded and nearly 4 million posts are posted on Facebook \cite{allen2017happens}, with this number increasing every year. Finding right information through this pile of information in short span of time is very difficult for user and is like finding needle in a haystack. Advancements in technologies have led to upswing of recommender systems as a solution to this problem.

Recommender systems are tools that use stored information to find recommended items that user might be interested in. Word 'item' is a generic term that can represent any type of content being recommended to the users. Apparels, movies, books, electronics, and computer systems all come under the word `item'. However, generally only one type of content is focused by a recommender system. Recommended items are calculated through a prediction function based on factors like past behavior, relations to other users, items similarity. This function calculates and predict user's probability to like a certain item. The system uses different information to learn behavior of a user through similar items, similar users, browsing history, or purchasing behavior to tailor recommendations. 

Many modern websites now use some form of recommender system to filter information and recommend items. For example, Facebook uses interaction behavior of user to arrange posts on news feed. Post from most interacted friends, pages, and groups are displayed on the top of news feed. Similarly, YouTube uses past browsing and watch history of user to recommend related videos. Almost every ecommerce website likes of Amazon uses recommendation algorithm to display list of items that are bought together. It can be said without a doubt that need of recommender systems is more than ever to match speed of data generation. Every recommender system has two important phases:
\begin{enumerate}
	\item \textbf{Learning Phase}: In this phase, behavior of user or relationship between other similar users or items is learned by the system to build a model that represents taste of user or relationship between items.
	\item \textbf{Decision Phase}: In this phase, model created in learning phase is used with preferences and constraints set by user to predict most suitable items for the user.
\end{enumerate}

The core of any recommender system is the learning phase which can be associated with data mining problem. Data mining is a process which uses different methods to extract valuable and interesting information from huge set of existing data. Well-known methods used in implementation of recommender system with data minding are association rule learning, classification, clustering, statistics, and machine learning algorithms \cite{amatriain2014recommender}. 

\subsection{Approaches for Recommender Systems}
\label{approachesForRS}
Use of recommendation system is not a new concept and over the years many different approaches have been proposed and implemented to achieve better results. Most of these approaches can be divided into two categories: \textit{traditional approaches} and \textit{advanced approaches}. Traditional approaches include content-based filtering, collaborative filtering, demographic, and hybrid techniques. These approaches are being used in practice for many years and are proven successful in most of the cases. Advanced approaches include learning rank and deep learning (\textit{Artificial Neural Networks}) and use latest research developments to implement more sophisticated recommender systems. Each approach has its own effectiveness in handling certain application domains and issues. Detailed explanation of each approach is left for another study, therefore, only collaborative filtering technique is discussed briefly as it is used in design of proposed recommender system.

\subsubsection{Collaborative Filtering}
\label{CF}
Collaborative filtering (CF) is a technique used to predict preference of user or rating of item based on other users' decisions and his own previous preferences. This approach assumes (1) User's preferences are not changed much over the time and (2) If two users A and B share the same review on one item, then B is likely to have review of A on different item that B has not encountered but A has. CF is one of the most widely implemented and well-known technique in recommender systems \cite{ricci2011introduction}. Amazon \cite{sarwar2001item} and Netflix \cite{wikiNetflix} both have collaborative filtering algorithms applied in their recommendation engines. 

Core of collaborative filtering technique is computation of similarity between items or users. For this purpose many algorithms are available such as Euclidean distance, Cosine Similarity, and Pearson correlation coefficient. CF approach can be further divided into three categories based on subject:
\begin{itemize}
	\item \textbf{User-based:} Similarity computation is used to find similar users to target user and items liked by them are recommended.
	\item \textbf{Item-based:} Similarity computation is used to find similar items to the one target user has liked in past and recommend those items.
	\item \textbf{Model-based:} Machine learning algorithms are used to develop a model for prediction of preferences of target user. 
	
\end{itemize}

\section{Design of VU-CRS}
\label{designAndImplementation}
This chapter presents data and design details of the Virtual University Course Recommender System (VU-CRS) using user-based collaborative filtering approach. The VU-CRS is designed and developed in the form of web-based project using PHP as back-end programming language. Front end is designed using HTML and CSS.

\subsection{Data}
\label{data}
The required data of VU-CRS included different courses offered at Virtual University and information of current students and past students for at least 4 --- 5 years. Since, students' information can have private data so getting required data through registrar office of University was not possible. Therefore, course data was fetched through main website of university. Study scheme page provides list of programs being offered and their respective list of courses for each semester. Link of study scheme page of each program was passed to custom parser to extract \textit{course code}, \textit{course title} and \textit{course type} (required or elective). This provided us with a raw data file consists of semester-wise courses detail for each degree. Using raw data file a complete list of courses was generated in a CSV file. There were few defects in the raw data that were fixed post parsing using MS Excel. First and foremost, there were a lot of duplicate courses which were removed. There were few courses with different course codes and same course titles. On deep investigation, it was found that some courses with same title were part of both BS and MS programs. Therefore, courses with same title were not updated or removed from the catalogue. Table \ref{porOfCdata} presents a portion of course catalogue. 

\begin{table}
	\centering
	\caption{Portion of Course Data}
	\label{porOfCdata}
	\begin{tabular}{|c|c|c|}
		\hline 
		\textbf{Course Code} & \textbf{Course Title} & \textbf{Type} \\ 
		\hline 
		ACC311	& Fundamentals of Auditing &	Required \\ 
		\hline 
		ACC501	& 	Business Finance	& 	Required \\ 
		\hline 
		BIF401	& 	Bioinformatics I	& 	Required \\ 
		\hline 
		BIF402	& 	Ethical and Legal Issues in Bioinformatics	& 	Required \\ 
		\hline 
		BIF501	& 	Bioinformatics II	& 	Required \\ 
		\hline 
		BIF601	& 	Bioinformatics Computing I	& 	Required\\ 
		\hline 
		BIF602	& 	Bioinformatics Computing II		& Required \\ 
		\hline 
		BIF604	& 	Special Topics in Bioinformatics	& 	Required \\ 
		\hline 
		BIF619	& 	Research Project	& 	Required \\ 
		\hline 
		BIO101	& 	Basic I-Biology		& Elective \\ 
		\hline 
	\end{tabular}
\end{table} 

For students' data, it was simulated through a custom built Java program which used raw data file to create a list of 40 students for each degree program. A random score between 40 and 99 was assigned for each course. Each student was assigned a sequential \textit{ID}, \textit{degree}, and \textit{number of semesters} currently studied which was between 1 to maximum semester in program. Simulated data does not reflect similarity with actual data but it served the purpose of testing our system. Table \ref{porOfSdata} presents a portion of students' studied courses.
\begin{table}
	\centering
	\caption{Portion of Student Data}
	\label{porOfSdata}
	\begin{tabular}{|c|c|c|c|}
		\hline 
		\textbf{Student ID} & \textbf{Course Code} & \textbf{Marks} & \textbf{Degree} \\ 
		\hline 
		1001	& 	PSY101	&	49	&	M.Sc. Applied Psychology \\ 
		\hline 
		1001	&	PSY404	&	50	&	M.Sc. Applied Psychology \\ 
		\hline 
		1001	&	PSY405	&	41	&	M.Sc. Applied Psychology \\ 
		\hline 
		1001	&	PSY502	&	79	&	M.Sc. Applied Psychology \\ 
		\hline 
		1001	&	STA630	&	67	&	M.Sc. Applied Psychology \\ 
		\hline 
		1001	&	PSY402	&	79	&	M.Sc. Applied Psychology \\ 
		\hline 
		1001	&	PSY403	&	64	&	M.Sc. Applied Psychology \\ 
		\hline 
		1001	&	PSY504	&	63	&	M.Sc. Applied Psychology \\ 
		\hline 
		1001	&	PSY610	&	49	&	M.Sc. Applied Psychology \\ 
		\hline 
		1001	&	PSY631	&	90	&	M.Sc. Applied Psychology \\ 
		\hline 
	\end{tabular}
\end{table} 

In the end, there were 2600 students and 470 courses in our data. From this point forward, \textit{Sdata}, \textit{Cdata}, and \textit{SCdata} are used to refer \textit{students}, \textit{courses}, and \textit{studied courses} data, respectively.

\subsection{Cold Start}
\label{coldstart}
The situation when there is not enough data present in system to calculate recommendations is called \textit{cold start}. In our case, when student has just registered to the system and has not entered any studied courses information leads to cold start. In order to handle this situation, feature of popular and top courses are provided in the system. Top and popular courses use na\"ive approach and display top 20 courses which are studied by most students and which have better average marks. Table \ref{popularCourses} presents list of 20 most popular courses taken by students. It is to be noted that top 4 courses are all required and to be studied by most students in first two semesters. One of Islamic studies or Ethics is also required to study. In general, all courses in this list are those which are offered in most degrees and are required (compulsory). Table \ref{topCourses} presents list of top 20 courses which yield highest average marks. Since, marks stored in the database are simulated, therefore, this list does not represent courses that actually produce highest marks.
\begin{table}
	\centering
	\caption{20 Popular Courses}
	\label{popularCourses}
	\begin{tabular}{|m{0.15\columnwidth}|m{0.40\columnwidth}|m{0.15\columnwidth}|m{0.10\columnwidth}|}
		\hline 		
		 \textbf{Course Code} & \textbf{Course Title} & \textbf{Course Type} & \textbf{Students} \\ 
		\hline 
			CS101 &	Introduction to Computing &	Required &	1806 \\ 
		\hline 
			ENG101 &	English Comprehension &	Required &	1400 \\ 
		\hline 
			ENG201 &	Business and Technical English Writing &	Required &	1307 \\ 
		\hline 
			PAK301 &	Pakistan Studies &	Required &	1145 \\ 
		\hline 
			ISL201 &	Islamic Studies &	Elective &	923 \\ 
		\hline 
			ETH201 &	Ethics (for Non-Muslims) &	Elective &	896 \\ 
		\hline 
			SOC101 &	Introduction to Sociology &	Required &	875 \\ 
		\hline 
			ECO401 &	Economics &	Required	 & 819 \\ 
		\hline 
			MGT211 &	Introduction To Business &	Required &	801 \\ 
		\hline 
			MGT101 &	Financial Accounting &	Required &	798 \\ 
		\hline 
			MTH302 &	Business Mathematics \& Statistics & Required & 792 \\ 
		\hline 
			MCM301 &	Communication skills &	Required &	695 \\ 
		\hline 
			STA301 &	Statistics and Probability &	Required &	621 \\ 
		\hline 
			MGT503 &	Principles of Management &	Required &	605 \\ 
		\hline 
			MGT301 &	Principles of Marketing &	Required &	574 \\ 
		\hline 
			CS201 &	Introduction to Programming &	Required &	543 \\ 
		\hline 
			PSY101 &	Introduction to Psychology &	Required &	500 \\ 
		\hline 
			ENG301 &	Business Communication &	Required &	485 \\ 
		\hline 
			STA630 &	Research Methods &	Elective &	478 \\ 
		\hline 
			MGT501 &	Human Resource Management &	Required &	452 \\ 
		\hline 
	\end{tabular}
\end{table} 

\begin{table}
	\centering
	\caption{20 Top Courses}
	\label{topCourses}
	\begin{tabular}{|m{0.15\columnwidth}|m{0.40\columnwidth}|m{0.15\columnwidth}|m{0.10\columnwidth}|}
		\hline 		
		\textbf{Course Code} & \textbf{Course Title} & \textbf{Course Type} & \textbf{Average Marks} \\ 
		\hline 
		ECTD520 & Teaching Practice (Long Term) & Required & 86 \\
		\hline
		MKT630 & International Marketing & Elective & 79 \\
		\hline
		BIF604 & Special Topics in Bioinformatics & Required & 78 \\
		\hline
		MTH645 & Fuzzy Logic and Applications & Required & 78 \\
		\hline
		FINI620 & Internship Report-Finance & Elective & 77 \\
		\hline
		BNKI620 & Internship Report-Banking & Elective & 77 \\
		\hline
		BT604 & Industrial Biotechnology & Required & 76 \\
		\hline
		FIN620 & Final Project-Finance & Elective & 76 \\
		\hline
		MKT620 & Final Project-Marketing & Elective & 76 \\
		\hline
		EDUA602 & Leadership and Management & Required & 75 \\
		\hline
		BNK620 & Final Project-Banking & Elective & 75 \\
		\hline
		CS721 & Network Performance Evaluation & Elective & 75 \\
		\hline
		CS620 & Modelling and Simulation & Required & 74 \\
		\hline
		BT503 & Environment Biotechnology & Required & 74 \\
		\hline
		CS718 & Wireless Networks & Elective & 74 \\
		\hline
		ELT620 & Thesis & Required & 74 \\
		\hline
		BT619 & Research Project & Elective & 74 \\
		\hline
		ECO613 & Globalization and Economics & Elective & 74 \\
		\hline
		EDU508 & Teaching of English Language & Required & 73 \\
		\hline
		MCM619 & Final Project-Mass Communication & Required & 73 \\
		\hline
	\end{tabular}
\end{table} 

\subsection{Flow Chart of VU-CRS}
\label{approachForVUCRS}
\begin{figure*}[htbp]
	\centerline{\includegraphics [width=0.97\textwidth]{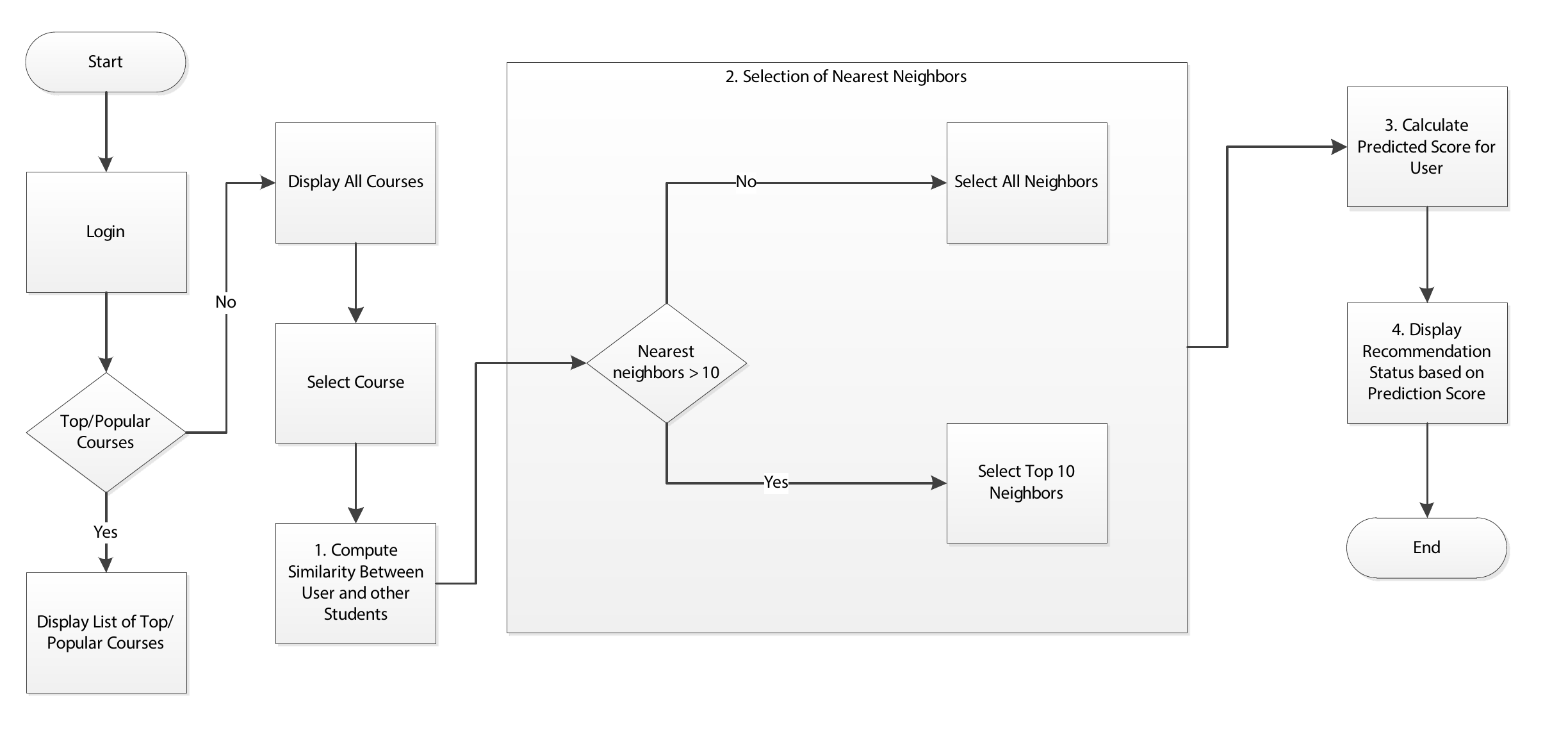}}
	\caption{Flow Chart of VU-CRS}
	\label{flowchart}
\end{figure*}

Cold Start issue discussed in previous section(\ref{coldstart}) helped to extract top taken courses and most popular courses. For any other case where there is at least one other student who has studied target subject can be processed for recommendation through system. complete flow chart of VU-CRS is given in figure \ref{flowchart}. 

various steps involved in calculation of recommendation status for a specific course are:
\begin{enumerate}
	\item Similarity Computation
	\item Selection of K nearest neighbors
	\item Calculation of Prediction score
	\item Displaying Recommendation Status
\end{enumerate}
Detail of each step is given in next section.

\subsection{Implementation}
\label{implementation}
Section \ref{RelatedWork} and Section \ref{RS} provided a brief overview of similar work done by researchers and approaches to implement recommender systems. Using reviewed work as an inspiration, user-based collaborative filtering approach is applied to implement VU-CRS. Complete algorithm is given in \ref{algorithm}.

\subsubsection{Similarity Computation}
First step in our system is to calculate similarity between target student and all other students who are present in the database. Computation of similarity depends on type of computational method used. As described in \textit{Collaborative Filtering} section \ref{CF}, there are multiple formulas available for similarity computation. There are 2600 students and 470 courses where one student cannot study more than 50 courses even for BS programs which makes data sparse, therefore, \textit{Cosine Similarity} method is used for VU-CRS. Equation \ref{eqCos} represents Cosine Similarity formula. To keep calculation overhead small, only those students who have studied target course are used to calculate similarity. This reduced the calculation to less than 15\% as one subject is not offered in every degree program.
\begin{equation}
\label{eqCos}
\cos ({\bf t},{\bf e})= {{\bf t} {\bf e} \over \|{\bf t}\| \|{\bf e}\|} = \frac{ \sum_{i=1}^{n}{{\bf t}_i{\bf e}_i} }{ \sqrt{\sum_{i=1}^{n}{({\bf t}_i)^2}} \sqrt{\sum_{i=1}^{n}{({\bf e}_i)^2}} }
\end{equation}

\subsubsection{Selection of Nearest Neighbors}
Equation \ref{eqCos} provided a list of similar students. The second step is to select nearest neighbors. In other words, who are more closely related to target students than the others. For this purpose, it is checked if number of similar students is greater than 10, then only top 10 are selected while others are discarded. However, if number is less than 10, then all of them are selected and passed to next phase.

\subsubsection{Prediction Score Calculation}
Prediction of score is the third step which uses past marks of target student and weighted average of nearest neighbors to calculate value which represents predicted marks of student in selected course. Prediction (rating) formula used for this purpose is given in \ref{eqPredictionScore}.  

\begin{equation}
\label{eqPredictionScore}
\mathrm{Rating}(\mathit{A,X})=  \overline{{r}_A} + {\sum_{i=1}^{n}{{sim(A,{B}_i) \times  (r_{(B_{i},X)} - \overline{r_{B_{i}}})}} \over \sum_{i=1}^{n}{{sim(A,{B}_i)}}} 
\end{equation}

Where \textit{X} is subject selected for recommendation status by student \textit{A}. Total Weighted bias is sum of weighted bias of all neighbors divided by sum of similarity score of all neighbors. Total weighted bias is added to average rating (marks) of target student \textit{A} to get final prediction value (marks) of \textit{A} in selected subject \textit{X}.

\subsubsection{Recommendation Status}
Once final prediction value is obtained from equation \ref{eqPredictionScore}, it is checked against set of conditions to find final recommendation status which is based on range of marks in percentage for grading scheme used at Virtual University \cite{vugrading}. The set of recommendation status is given in table \ref{recommendationStatus}.
\begin{table}
	\centering
	\caption{Recommendation Status}
	\label{recommendationStatus}
	\begin{tabular}{|c|c|c|}
		\hline 
		\textbf{Prediction Value} & \textbf{Grade} & \textbf{Status} \\ 
		\hline 
		-1 & N-A & Already Studied \\
		\hline 
		-2 & N-A & Not Enough Data \\
		\hline 
		$\geq$90 & A+ & High Recommendation \\
		\hline 
		$\geq$85 & A & High Recommendation \\
		\hline 
		$\geq$80 & A- & High Recommendation \\
		\hline 
		$\geq$75 & B+ & Recommended \\
		\hline 
		$\geq$71 & B & Recommended \\
		\hline 
		$\geq$68 & B- & Recommended \\
		\hline 
		$\geq$61 & C & Low Recommendation \\
		\hline 
		$\geq$50 & D & Student Decision \\
		\hline 
		$<$50 & F & Not Recommended \\
		\hline 
	\end{tabular}
\end{table} 

-1 prediction value represents student has already studied the selected subject whereas -2 represents there is very low number of students (currently set at 3) who have studied the subject therefore system does not have enough data to make any recommendation. Remaining values are expected Grades, where value $\geq$50 represents \textit{D} grade which is equivalent to very low grade point average. Selection of course in such case might not be good for fulfillment of minimum CGPA degree requirement, therefore, decision has been left with student.

\subsubsection{Algorithm}
\label{algorithm}
Complete algorithm based on all steps defined in previous sub-sections is given in \ref{algorithm}.
\begin{algorithm}
	\DontPrintSemicolon
	\KwData{$VUCRS(targetID, X)$}
	
	\Begin{
		
		A = FetchTargetStudentFromDB(targetID)\;
		\If{targetHasStudied(A, X)} {
			\Return{-1}
		}
		\tcc{Students Who Have Studied X}
		students = FetchAllStudentsFromDB(X)\;
		\ForEach{B in Students} {
			simArray = CosineSimilarity(A, B)\;
		}
		
		\If{sizeof($simArray$) $\leq$ 3} {
			\Return {-2}
		}
		
		\tcc{Select K-Nearest Neighbors}
		kNN = NearestNeigbors(simArray)
		
		\tcc{Calculate Prediction Rating}
		rating = RatingFunction(A, X, kNN)
		
		\Return{rating}
	}
	\caption{VU-CRS Algorithm}
	\label{VUCRSAlgo}
\end{algorithm}

\section{Testing and Evaluation}
\label{testing}
Implemented VU-CRS was evaluated against various students and courses. Different courses for one student were first tested and then one course against different set of students was tested to get complete information. Table \ref{testCases} provides a complete testing results. 

Test case number 1 to test case number 5 represent recommendation of five different courses for one student of \textit{M.Sc. Applied Psychology}. It can be seen that his average marks are $69$ which heavily influence predicted marks, because in the prediction formula (eq \ref{eqPredictionScore}) target student's average marks are added in final prediction score. However, final prediction score also depends on total weighted bias of similar students. Recommendation status for this student are either \textit{recommended} or \textit{low recommendation}, as 3 out of 5 expected grades are \textit{C}. 

Test case number 6 to test case number 10 represent recommendation of five courses for a different student of \textit{BS Software Engineering}. Courses selected for recommendation were mainly required courses. In comparison with previous student, average marks for this student are $83$ which can be seen from predicted marks that most of the values in that column against this student are in higher grades (i.e. \textit{A} or \textit{A-}). Interesting point here is that difference between course average marks and student's average marks is as high as $19$, which clearly suggest that expected marks for any new course are influenced by previous marks. One course \textit{CS101} was already studied by student, so no predicted marks were calculated and no expected grade was possible.

Remaining cases tested different strategy by fixing one course for five student of same degree. First it was \textit{CS304 - Object Oriented Programming} course with $71$ average marks. Expected grade for each student was different because difference between each student's average marks and course average marks was varying. But for other set of tests, where course code \textit{MGT611} was used, average marks for this course were 66 which were not very much different from average marks of five selected students. And it can be seen that 3 students were give \textit{Low Recommendation} status and 2 with \textit{Recommended} status.

\begin{table*}[t]
	\centering
	\caption{Test Cases}
	\label{testCases}
	\begin{tabular}{|m{0.05\linewidth}|m{0.05\linewidth}|m{0.1\linewidth}|m{0.05\linewidth}|m{0.15\linewidth}|m{0.05\linewidth}|m{0.05\linewidth}|m{0.05\linewidth}|m{0.05\linewidth}|m{0.1\linewidth}|}
		\hline 
		\textbf{Test Case No} &	\textbf{Student ID}	& \textbf{Degree} &	\textbf{Course Code} &	\textbf{Course Title} &	\textbf{Course Average}	& \textbf{Student Average} &	\textbf{Predicted Marks} &	\textbf{Expected Grade} &	\textbf{Status} \\
		\hline 
		1 & 1039 & M.Sc. Applied Psychology & CS101 & Introduction to Computing & 67 & 69 & 65 & C & Low Recommendation \\
		\hline
		2 & 1039 & M.Sc. Applied Psychology & MKT630  &  International Marketing & 79 & 69 & 76 & B+ & Recommended \\
		\hline
		3 & 1039 & M.Sc. Applied Psychology & PSY632 & Theory \& Practice of Counseling & 67 & 69 & 63 & C & Low Recommendation \\
		\hline
		4 & 1039 & M.Sc. Applied Psychology & PSY610 & Neurological Bases of Behavior & 66 & 69 & 71 & B & Recommended \\
		\hline
		5 & 1039 & M.Sc. Applied Psychology & MCM431 & Development Communication & 70 & 69 & 67 & C & Low Recommendation \\
		\hline
		6 & 1920 & BS Software Engineering & CS302 & Digital Logic Design & 67 & 83 & 86 & A & High Recommendation \\
		\hline
		7 & 1920 & BS Software Engineering & CS304 & Object Oriented Programming & 71 & 83 & 83 & A- & High Recommendation \\
		\hline
		8 & 1920 & BS Software Engineering & PHY301 & Circuit Theory & 68 & 83 & 84 & A- & High Recommendation \\
		\hline
		9 & 1920 & BS Software Engineering & MCM301 & Communication skills & 69 & 83 & 76 & B+ & Recommended \\
		\hline
		10 & 1920 & BS Software Engineering & CS101 & Introduction to Computing & 67 & 83 & N-A & N-A & Already Studied \\
		\hline
		11 & 1876 & BS Information Technology & CS304 & Object Oriented Programming & 71 & 67 & 70 & B- & Recommended \\
		\hline
		12 & 1877 & BS Information Technology & CS304 & Object Oriented Programming & 71 & 84 & 86 & A & High Recommendation \\
		\hline
		13 & 1878 & BS Information Technology & CS304 & Object Oriented Programming & 71 & 74 & 72 & B & Recommended \\
		\hline
		14 & 1879 & BS Information Technology & CS304 & Object Oriented Programming & 71 & 58 & 60 & D & Student Decision \\
		\hline
		15 & 1880 & BS Information Technology & CS304 & Object Oriented Programming & 71 & 70 & 66 & C & Low Recommendation \\
		\hline
		16 & 2786 & BS (Business Administration) & MGT611 & Business \& Labor Law & 66 & 65 & 64 & C & Low Recommendation \\
		\hline
		17 & 2787 & BS (Business Administration) & MGT611 & Business \& Labor Law & 66 & 67 & 69 & B- & Recommended \\
		\hline
		18 & 2788 & BS (Business Administration) & MGT611 & Business \& Labor Law & 66 & 71 & 75 & B+ & Recommended \\
		\hline
		19 & 2789 & BS (Business Administration) & MGT611 & Business \& Labor Law & 66 & 64 & 61 & C & Low Recommendation \\
		\hline
		20 & 2790 & BS (Business Administration) & MGT611 & Business \& Labor Law & 66 & 65 & 64 & C & Low Recommendation \\
		\hline
	
	\end{tabular}
\end{table*} 

\subsection{Evaluation}
Evaluation of recommender system is very important and considered necessity in many cases. A well-defined recommender system must be evaluated against different available metrics. A self-design metric can sometimes provide better feedback but well-established metrics are generally enough to get convincing feedback. Measuring accuracy of system through different metrics can yield different results as each has its own effectiveness. Two of the most effective metrics used to measure accuracy of a rating prediction based recommender systems are \textbf{Mean Absolute Error} and \textbf{Root Mean Squared Error}. Both provide measure of how predicted value is different from actual value. For evaluation of VU-CRS, Mean Absolute Error (given in eq-\ref{eqMAE}) is used.

\begin{equation}
\label{eqMAE}
MAE = (\frac{1}{n})\sum_{i=1}^{n}\left | y_{i} - x_{i} \right |
\end{equation}

Where $ y_{i} $ represents predicted value for $ i_{th} $ course and $ x_{i} $ represents actual value for same course.

Fifty different students with two courses for each student with known marks were used for evaluation purpose. It was observed that difference between predicted value and actual value remained less than 10\% for most observations where for very few ($<$10\% observations) difference fluctuated between $14\%$ and $20\%$. However, overall Mean Absolute Error calculated through eq-\ref{eqMAE} came out to be $5.12$ for 100 observations. Which is acceptable error because percentage range for different grades is 5 marks for most of the grades.

\section{Conclusion and Future Work}
\label{conclusion}

There are number of both public and private Universities in Pakistan providing education at Undergraduate, Postgraduate, and Doctorate level with average annual enrollment in any such institution in thousands. Virtual University of Pakistan is country's first university providing education in digital programs through distance-learning platform. Virtual University offers education in many different majors and various areas of study are available. Multiple courses are offered in each field of study that satisfy several general requirements of degree in that area. Course selection process is quick and easy for those students who had already planned out their courses but for some students it can become difficult to select courses which align with their competency and are relatively easier to earn good grade. Selection of courses which are outside student's interest and are naturally considered difficult to get good grades (as per past records) can impact overall percentage and CGPA of student. Therefore, keeping this issue in mind, a course recommender system for Virtual University has been developed with the aim to help students in course selection process. A brief overview of similar work done by researchers was presented along with different approaches that can be used to implement a recommender system are discussed. Using reviewed work as an inspiration, user-based collaborative filtering approach was applied to implement VU-CRS. Required data (courses and students information) for testing purpose was extracted from official website of VU. A dataset of 470 courses and 2600 students was prepared. Students' data was simulated using custom-built Java program. Testing of implemented system included various scenarios such as fixing one student and finding recommendation status for different courses and fixing one course and finding recommendation status for different students. Results from both kinds of tests showed that predicted marks are heavily dependent on student's average marks and average marks obtained by similar students in that particular course. Accuracy of results was measured using Mean Absolute Error with testing predicted values for 100 observations with known actual values. MAE value came out to be nearly 5\% which was acceptable value looking at grading scheme followed by VU. 

Data of students was simulated for testing purpose, with the availability of actual data, more accurate test results can be generated. Approach used in this implementation uses past marks only. With the availability of more data fields, different approaches for similarity computation can be used. Cold start issue discussed in this approach followed a na\"ive approach by displaying top/popular courses. In future, a psychological test based approach can be used to generate more personalized results. Processing language used for implementation purpose is PHP, which can slow down the entire process when number of students can get large. Other languages such as Python for computation purposes using frameworks like Django can improve processing speed.

\bibliographystyle{IEEEtran}


\bibliography{vucrsbib}

\end{document}